\documentclass[journal]{IEEEtran}
\usepackage{amsmath,amsfonts}
\usepackage{algorithm}
\usepackage{array}
\usepackage[caption=false,font=normalsize,labelfont=sf,textfont=sf]{subfig}
\usepackage{textcomp}
\usepackage{stfloats}
\usepackage{url}
\usepackage{verbatim}
\usepackage{graphicx}
\usepackage{cite}

\newcommand{\cmark}{\ding{51}}%
\newcommand{\xmark}{\ding{55}}%

\usepackage{graphicx}
\usepackage{subcaption}
\usepackage{amsmath}
\usepackage{amssymb}
\usepackage{booktabs}
\usepackage{makecell}
 
\usepackage{multirow}
\newcolumntype{x}[1]{>{\centering\let\newline\\\arraybackslash\hspace{0pt}}p{#1}}
\usepackage{color, colortbl}
\usepackage{xcolor}
\makeatletter
\DeclareRobustCommand{\itshape}{%
  \not@math@alphabet\itshape\mathit
  \fontshape\itdefault\selectfont
  \color{gray}%
}
\makeatother
\usepackage{algorithm}
\usepackage{algpseudocode}
\usepackage{pifont}
\usepackage{listings}
\usepackage{etoolbox}
\usepackage{lipsum}
\usepackage{mathtools}
\usepackage{verbatim}

\usepackage[pagebackref,breaklinks,colorlinks]{hyperref}
\definecolor{commentgreen}{rgb}{0.0, 0.27, 0.13}

\usepackage{natbib}
\setcitestyle{numbers}
\setcitestyle{square}
\setcitestyle{citesep={,}}
\usepackage{hyperref}
\usepackage{url}

\definecolor{darkgreen}{rgb}{0.0, 0.4, 0.13}

\begin{document}

\title{ASiT: Local-Global Audio Spectrogram vIsion Transformer for Event Classification}

\author{Sara Atito,~\IEEEmembership{Member,~IEEE,}
 \quad Muhammad Awais,~\IEEEmembership{Member,~IEEE,} \quad Wenwu Wang,~\IEEEmembership{Senior Member,~IEEE,} \\
 \qquad Mark D Plumbley,~\IEEEmembership{Fellow,~IEEE,} \qquad Josef Kittler,~\IEEEmembership{Life Member,~IEEE,}

\thanks{Centre for Vision, Speech and Signal Processing (CVSSP) and Surrey Institute for People-Centred AI,  University of Surrey, Guildford, UK.}
      
}



\maketitle

\begin{abstract}

Transformers, which were originally developed for natural language processing, have recently generated significant interest in the computer vision and audio communities due to their flexibility in learning long-range relationships. 
Constrained by the data hungry nature of transformers and the limited amount of labelled data, most transformer-based models for audio tasks are finetuned from ImageNet pretrained models, despite the huge gap between the domain of natural images and audio.
This has motivated the research in self-supervised pretraining of audio transformers, which reduces the dependency on large amounts of labeled data and focuses on extracting concise representations of audio spectrograms.
In this paper, we propose  \textbf{L}ocal-\textbf{G}lobal \textbf{A}udio \textbf{S}pectrogram v\textbf{I}sion \textbf{T}ransformer, namely ASiT, a novel self-supervised learning framework that captures local and global contextual information by employing group masked model learning and self-distillation.  
We evaluate our pretrained models on both audio and speech classification tasks, including audio event classification, keyword spotting, and speaker identification. We further conduct comprehensive ablation studies, including evaluations of different pretraining strategies.
The proposed ASiT framework significantly boosts the performance on all tasks and sets a new state-of-the-art performance in five audio and speech classification tasks, outperforming recent methods, including the approaches that use additional datasets for pretraining.

\end{abstract}


\section{Introduction}
\label{sec:intro}

\IEEEPARstart{T}{hroughout} the history of pattern recognition and machine learning, there are numerous examples where the learning methodology developed for one dimensional time varying signals or data have been shown to be equally relevant to two dimensional signal domains and vice versa. The instances include the use of factor analysis in automatic speech recognition, proposed to deal with different speech environments back in 2000 \cite{Saul_sap2000}, successfully extended to 2D for coping with different poses in face recognition in \cite{Prince_pami2008}. More recent is the application of a new deep neural network (DNN) architecture, the transformer~\cite{vaswani2017attention},  proposed for natural language processing, to the 2D domain of image processing \cite{dosovitskiy2020image}. In this paper, we investigate the reverse task, namely the applicability of techniques developed for 2D image analysis, to a 1D signal domain. In particular, we consider the relevance of transformer based image classification approaches to the problem of audio classification.  

Although audio is a 1D signal, working with its spectral representation partly bridges the gap between the 1D and 2D domains. However, the frequency/time nature of a spectrogram differs significantly from a conventional image, where the statistical properties (pixel relationships) of the signal in any arbitrary direction and its physical meaning are the same. This contrasts with the spectrogram, where the image axes represent different physical phenomena, namely, time and frequency. Moreover, in a spectrogram, the classes are superimposed, whereas in a scene image containing several objects, the classes are adjacent to each other. These differences are significant enough to raise doubts about the direct applicability of the techniques developed for image analysis to the audio classification problem. These doubts are addressed in the case of convolutional neural networks (CNNs) \cite{hershey2017cnn,kong2020panns}. However, the applicability of the attention-based model is still an open question, with a few works~\cite{gong2021ast,gong2022ssast,baade2022mae,chong2023masked} validating the use of vision transformers (ViT) for audio.

We shall confine our study to ViTs, which are gaining rapid popularity. ViTs are very data hungry~\cite{dosovitskiy2020image,atito2021sit}, and their training is computationally costly. This often necessitates using existing pretrained models for developing new applications, spanning different user domains. The most popular are models pretrained on the ImageNet \cite{krizhevsky2012imagenet}, which are the models of choice, irrespective of their relevance to the application concerned. However, ImageNet pretraining is clearly not ideal for decision making, e.g. in medical domain~\cite{atito2022sb}. This data hungry nature of ViTs has recently been alleviated by group masked model learning (GMML) based self-supervised learning (SSL) methods, introduced in SiT~\cite{atito2021sit}. At its core, GMML is using the principle of denoising/masked autoencoder~\cite{vincent2008extracting}. GMML randomly masks various regions within the input image, potentially covering groups of connected patches and leaves groups of connected patches unmasked. The goal of learning is to recover missing information in the masked regions from visible information (context). Contrastive learning~\cite{vandenOord2018,wu2018unsupervised,hjelm2018learning,chen2020simple,NEURIPS2020_29539ed9,grill2020bootstrap} is another line of SSL approaches that aim to maximise the similarity between two augmented views of the same image and maximise the disagreement between different images. These approaches use various techniques, e.g. large batch size, memory banks, asymmetric projection heads, etc, to boost the performance and more importantly to avoid representation collapse. 

Thanks to the recent advance in the SSL approaches, the self-supervised pretraining of DNNs, without using labelled data, for the first time, outperformed supervised pretraining of DNNs in multiple computer vision downstream tasks like classification, segmentation etc. This enabled the use of domain specific pretraining of DNNs for high performance on several downstream tasks without using labelled data. 

The key questions addressed in this paper are: i) can masked image modelling successfully be applied to audio data, given the inherent differences between a 2D image and a spectrogram, and ii) how should the pretraining methodology be adapted to gain from SSL as much as possible. The first question has already been partially answered in earlier studies
~\cite{gong2021ast,gong2022ssast,baade2022mae,chong2023masked,chen2022beats} and we confirm here that SSL can be applied effectively to the audio classification tasks. However, a straightforward application of SSL is not enough. The use of masked autoencoder does not guarantee that sufficient discriminatory information is retained by the pretrained model \cite{baade2022mae} as further discussed in Section \ref{sec:related_works}.

While reconstruction is an essential ingredient, in order to preserve discriminatory information, an SSL method should promote similarity learning. To this end, we propose a novel transformer learning framework, constituted by a  teacher-student architecture, which is trained using a distillation approach guided by a contrastive loss, and by reconstruction error produced by an integrated reconstruction head. We argue that the typical global similarity learning using a classification token is insufficient to capture the relevant information, and demonstrate that the proposed local similarity learning mechanism is also crucial to successful pretraining for multilabel audio classification tasks. More specifically, we propose a contrastive-loss-based learning method by distillation, using the alignment of local features derived for the original and masked spectrograms as a measure of similarity. The masked spectrogram reconstruction and the combination of global/local similarity learning  produces effective pretraining for downstream audio data processing tasks.

We evaluate the proposed SSL pretraining method on several benchmarking audio and speech classification datasets. We demonstrate the ability of the proposed method to learn the underlying properties of audio data using pretraining and finetuning purely on audio datasets, without resorting to auxiliary image data sets. In fact, we show that the performance achieved by training exclusively on audio datasets delivers the best results, which exceed the state-of-the-art (SOTA) by a significant margin. In summary, our contributions include:

\begin{itemize}
\item A local-global audio spectrogram vision transformer (ASiT) for enhanced audio representation.
\item A novel self-supervised pretraining method for ASiT, which combines masked spectrogram reconstruction with local-global similarity learning using distillation.
 \item Extensive evaluation of ASiT on benchmarking datasets, with the results that define a new state of the art performance, demonstrating the merits of the proposed methodology. 
\end{itemize}


\begin{figure*}[t!]
    \centering
    \includegraphics[width=1\textwidth]{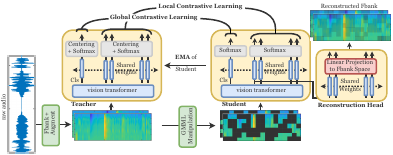}
    \caption{The proposed self-supervised framework (ASiT). For a given 10-second audio spectrogram, two random augmented views of 6-second each (clean spectrograms) are generated and fed to GMML based manipulation block to obtain the masked spectrograms. The clean and masked spectrograms are fed to the teacher and student networks, respectively. The recovery of the transformed information from the non-transformed class-token and data-tokens indicates that the network has learnt the semantics of the local as well as the global representation of the given audio and learnt useful inductive bias by learning local statistical correlation in the spectrogram.}
     \vspace{-0.2cm}
    \label{fig:arch}
\end{figure*}

\section{Related Works}
\label{sec:related_works}
Unsupervised learning offers advantages in improving the audio representation by leveraging large scale data without labels \cite{Xuyong2017}. SSL aims to learn audio representations with pairs of similar inputs derived from unlabelled data, by exploiting temporal consistency \cite{vandenOord2018} or data augmentation \cite{saeed2021contrastive, Liuxubo2021}. After its wide adoption in natural language processing (NLP) and computer vision, SSL is attracting increasing interest from audio community.
In early works, SSL has been applied mainly to speech \cite{chung2018speech2vec,Baevski2019, Baevski2020, BRiviere2020}. For example, Speech2Vec \cite{chung2018speech2vec}, which is inspired by Word2Vec \cite{mikolov2013efficient}, learns word embedding from acoustic features of speech, such as mel-frequency cepstral coefficients, using an RNN based encoder-decoder trained with skipgram or continuous bag-of-words approaches. 

SSL has been recently used to learn general purpose audio representations. For example, Audio2Vec \cite{tagliasacchi2020pre} learns a representation via reconstructing a spectrogram slice in terms of its preceding and following slices, using CNN based backbone and mel-spectrogram as input. 
In CLAR \cite{al2021clar} and COLA \cite{saeed2021contrastive}, which are built on the visual SSL method SimCLR \cite{chen2020simple}, the network is trained to maximize the representations of different views in latent space and reduce the reconstruction loss of the original waveform or spectrogram input, with different augmentations, such as random size cropping, Gaussian noise addition, and mix-back where the incoming patch is mixed with a background patch. These methods often assume that audio segments in temporal neighborhood have more similar representations, while those far away in time have more different representations. Nevertheless, this is not the case for repetitive sounds such as music and car sirens, where distant segments can have similar representations.

Different from the above methods, where training is performed by
contrast between positive and negative samples, the distillation-based approaches such as BYOL-A \cite{niizumi2021byol} and ATST \cite{li2022atst} learn general audio representations by minimizing the difference (e.g. in mean squared error) between the embeddings of the same input with the contrasts obtained via data augmentation. In this case, no negative samples are required for learning the representation. For example, in BYOL-A \cite{niizumi2021byol}, BYOL \cite{grill2020bootstrap} is applied to audio by learning the similarity between the two views with one learned from a randomly cropped single segment and the other from its augmented version obtained with e.g. mixup and random resized crop (RRC). Different from BYOL-A, in ATST \cite{li2022atst}, the two views were learned from two randomly cropped and then augmented mel-spectrogram segments with a teacher-student transformer architecture. The distillation-based methods have been reported to achieve state-of-the-art performance. However, this method can potentially lead to trivial solutions, i.e. collapsed representations, which require careful design of network architectures or training algorithms. For example, in Barlow Twins \cite{zbontar2021barlow}, the outputs of the network twins which take augmented samples as inputs are compared using cross-correlation, which promotes similarity between the learned embeddings from the network twins, while reducing the redundancies of the learned representations.

Inspired by the masked language modeling (MLM) \cite{devlin2018bert} and masked image modeling (MIM) \cite{atito2021sit}, several masked acoustic modeling (MAM) approaches such as SSAST \cite{gong2022ssast} and MAE-AST \cite{baade2022mae} were introduced which learn to reconstruct the masked time-frequency patches of an arbitrary shape from a given spectrogram. This offers potential advantages over the methods such as Audio2Vec \cite{tagliasacchi2020pre}, which focus on temporal modelling via reconstructing masked temporal spectrogram frames. In SSAST \cite{gong2022ssast}, the transformer based AST model \cite{gong2021ast} is pretrained on unlabelled audio from AudioSet and Librispeech with joint generative and discriminative masked spectrogram patch modeling. SSAST uses a very high masking ratio, hence the vast majority of self-attention is computed on masked patches. On the other hand, MAE-AST \cite{baade2022mae} and Audio-MAE \cite{xu2022masked} introduce the encoder-decoder architecture from masked autoencoder \cite{baade2022mae}, where a large encoder is used to learn on unmasked input while a smaller decoder is used to reconstruct masked input with encoder outputs and masked tokens.

Despite the promising performance of the transformer models in capturing local representations as in SSAST~\cite{gong2022ssast}, MAE-AST~\cite{baade2022mae}, and global representation as in ATST \cite{li2022atst}, the transformer architecture is limited in capturing optimal local contextual information or global relational information from audio. Modelling both local and global information optimally, nevertheless, can be crucial for detecting transient acoustic events such as gun shots as well as symphony. In this paper, we propose a novel self-supervised pretraining method for local context modelling via token similarity learning, and global representation learning via contrastive instance classification. Such combination allows the transformer model to capture fine-grained region/contextual dependencies as well as global representation learnt from data, rather than injecting any  inductive biases as in the case of CNNs. As a result, this novel framework significantly improves the quality of the learned audio representations.

\section{Methodology}
\label{sec:method}

In this section, we introduce ASiT, a self-supervised framework based on vision transformers for audio event representations. 
Similar to \cite{gong2021ast}, we employed log mel-spectrogram \cite{rabiner2010theory} as the input to the ViT instead of using the raw waveform directly. Spectrograms are used extensively in the fields of audio, speech, and music as they somewhat model human hearing perception and contain abundant low-level acoustic information, which has similar characteristics to images, making them more suitable for ViTs.

In Section \ref{sec:GMML}, we briefly summarise group masked model learning \cite{atito2022gmml} and the adoption of it to audio domain. Following, in Section \ref{sec:LG_rep}, we explain our proposed method of self-learning of data and class tokens conceptualisation, with the incorporation of knowledge distillation \cite{hinton2015distilling}.

\subsection{Group Masked Model Learning}
\label{sec:GMML}

Masked image modelling, which in principle is similar to the masked language modelling (MLM) used in BERT~\cite{devlin2018bert}, has been first 
proposed in SiT \cite{atito2021sit} and employed in several recent vision \cite{bao2021beit,atito2021mc,xie2022simmim} and audio \cite{gong2022ssast,baade2022mae,chong2023masked} works. 

The main idea of GMML is to mask various regions within the input image, potentially covering a group of connected patches representing a ``significant'' part of a given visual input and recover them by learning a model. The underlying hypothesis is that, by recovering the masked parts from the unmasked parts, based on the context from the whole visual field, the network will implicitly learn the notion of visual integrity. Intuitively the network will only be able to recover the missing information if it learns the characteristic properties of visual stimuli corresponding to specific actions impacting on the visual input. This way of masking strategy proved efficiency in image domain, especially when dealing with small datasets. It compels the model to incorporate an inductive bias, a critical component often absent in transformer models.

There are several ways to mask the various regions in the spectrogram. For example, replacing the randomly selected regions with zeros, noise, or with patches from other spectrograms. Note that the masked regions are manipulated randomly, i.e. the regions with audio or speech and the regions with noise or silence are manipulated with equal probability. Our hypothesis is that each pixel in a given log mel-spectrogram has a semantic meaning associated with it and the network is required to learn to reconstruct the information corresponding to both dominant (speech and audio) as well as non-dominant (noise and silence) concepts with equal importance, which equips the network in generalizing well for unseen data.

As for the masking strategy, the time and frequency are not treated differently during masking, i.e. we apply the random masking without any prior to allow the model to learn both the temporal and frequency structure of the data.

For spectrogram reconstruction, we propose to use the vision transformer as a group masked autoencoder, i.e. visual transformer autoencoder with group masked model learning. By analogy to autoencoders, our network is trained to reconstruct the input spectrogram through the output tokens of the transformer. 

A vision transformer receives as input a sequence of patches obtained by tokenising the input spectrogram $\mathbf{x} \in \mathbb{R}^{T \times F}$ into $n$ flattened $2D$ patches of size $p \times p$ pixels, where $T$ and $F$ are the time and frequency resolution of the input spectrogram and $n$ is the total number of patches. The $n$ patches are then flattened and fed to a linear projection to obtain $n$ tokens, which are then passed to the vision transformer. It's important to note that in the context of vision transformers, a patch refers to a small, square segment of the input image/spectrogram. After the initial linear embedding stage in a ViT, each patch embedding is referred to as a token.

To pretrain transformers as group masked autoencoder, the GMML manipulated spectrogram $\mathbf{\hat{x}}$ is first obtained by randomly replacing a significant part of the spectrogram with zeros or with patches from another spectrogram. Note that unlike in natural language processing, partial transformation of a patch is possible. Please refer to the ablation studies in Section \ref{sec:ablation} for details.
 
The reconstructed spectrogram $\mathbf{\bar{x}}$ is obtained by passing the masked spectrogram $\mathbf{\hat{x}}$ to the transformer encoder $E(\cdot)$ (the vision transformer in the student block in Figure \ref{fig:arch}) and feeding the output to a light decoder $D(\cdot)$ (the reconstruction head block in Figure \ref{fig:arch}). The decoder consists of 3 fully connected layers; the first two with 2048 neurons and GeLU \cite{hendrycks2016gaussian} non-linearity each, and the last bottleneck layer with 256 neurons, followed by a transposed convolution to return back to the spectrogram space. After the pretraining stage, the light decoder $D(\cdot)$ is dropped and only the encoder $E(\cdot)$ is used for the downstream task. The importance of using the representation before the nonlinear projection, i.e. decoder head, is due to loss of information induced by the reconstruction loss. Particularly, the decoder head generally learns task-specific features, i.e. spectrogram reconstruction, which might rescind information that may be useful for the downstream task.

For the spectrogram reconstruction task, we use the $\ell_1$-loss between the original and the reconstructed spectrogram as shown in Equation \ref{eq:l1-pixel}. We compute the loss only on the masked pixels, similar to \cite{devlin2018bert,atito2022gmml}.
\begin{equation}
\label{eq:l1-pixel}
\mathcal{L}(\mathbf{W})_{\rm recons} = \sum_k^N \left( \sum_i^T \sum_j^F M_{i,j}^k \times | x_{i,j}^k - \bar{x}^k_{i,j} | \right),
\end{equation}
\begin{equation}
    M_{i,j}^k = 
\begin{cases}
    1,              & \text{if } x_{i,j} \text{ in spectrogram } k \text{ is manipulated}\\
    0,              & \text{otherwise.}
\end{cases}
\end{equation}
where $\mathbf{W}$ denotes the parameters to be learned during training, $N$ is the batch size, and $\mathbf{M}$ is a binary matrix with 1 indicating the manipulated pixels.

\subsection{Local and Global Pseudo-Label Learning}
\label{sec:LG_rep}

Group masked model learning is the key component of ASiT as it implicitly learns the notion of visual integrity. This notion of visual integrity can be further enhanced by using pseudo labels that can be generated automatically, based on some attributes of the data. The term ``pseudo'' in this context refers to automatically generated labels derived from a teacher model’s output during knowledge distillation. The idea of assigning a local and global pseudo labels for the spectrogram is inspired from DINO \cite{caron2021emerging} framework. DINO focuses on assigning one global representation for the whole input. Unlike DINO, we propose to use self knowledge distillation to learn a pseudo label for each token in the spectrogram, as well as a global pseudo label for the whole spectrogram. Our hypothesis  is that modelling only one representation for the whole spectrogram can lead to sub-optimal representation learning. Thus, we investigate the role of learning the appropriate representation for each of the data tokens through knowledge distillation. 

In knowledge distillation, a student network ${s}^{\theta}(\cdot)$ is trained to match the output of a given teacher network ${t}^{\phi}(\cdot)$, where $\theta$ and $\phi$ are the parameters of the student and the teacher networks, respectively. In this work, we employ the same network architecture for both the student and the teacher, i.e. the teacher is a momentum encoder~\cite{he2020momentum} of the student, where the parameters of the teacher network are updated from the past iterations of the student network using exponential moving average of the student weights with the following update rule: $\phi \leftarrow \lambda \phi + (1 - \lambda) \theta$. 

In our case, the network architecture of the student comprises three components: A vision transformer backbone $s_{\rm b}(\cdot)$, followed by two projection heads attached to the output of the transformer. The first projection head ${s}_{\rm lcl}(\cdot)$ is for local contrastive learning and the second projection head  ${s}_{\rm gcl}(\cdot)$ is for global contrastive learning. The local contrastive learning projection head  $s_{\rm lcl}(\cdot)$ consists of 3 fully connected layers; the first two with $2048$ neurons and GeLU non-linearity each, and the last bottleneck layer with $256$ neurons. The output of the bottleneck layer is $\ell_2$ normalised and directly connected to a weight-normalised fully connected layer with $K=8192$ neurons. The global contrastive learning projection head $s_{\rm gcl}(\cdot)$  has a similar design to $s_{\rm lcl}(\cdot)$, where the output layer has $C=8192$ neurons. In our implementation, we shared the weights of the local and global contrastive learning projection heads. However, we expect further improvements by tuning $K$ and $C$ hyper-parameters as well as the choice of sharing, partially sharing, or not sharing the weights of the contrastive learning projections heads.  

In order to generate a local pseudo label for the input spectrogram $\mathbf{x}$, the clean spectrogram is fed to the backbone of the teacher network $t_{\rm b}(\mathbf{x})$ and the output of the data tokens  are passed to the local contrastive learning projection head to obtain $\mathbf{z}_t \in \mathbb{R}^{n \times K}$. As for the student network, the GMML-based manipulated spectrogram $\mathbf{\hat{x}}$ is passed to the student network to obtain $\mathbf{z}_s \in \mathbb{R}^{n \times K}$. The task is to match the output of the student network to the output of the teacher network, employing the Kullback-Leibler divergence (KL).

Training the student to match the teacher output can easily lead to a trivial constant (i.e. collapsed) embeddings. To avoid the model collapse, we adopted the centering and sharpening of the momentum teacher outputs introduced in~\cite{caron2021emerging}. The centering encourages the output to follow a uniform distribution (similar to label smoothing), where a center C is added to the teacher distribution. Meanwhile, the softmax function's temperature parameter effectively modulates the sharpness of the output probabilities. Applying both operations balances their effects, which is sufficient to avoid a collapse in the presence of a momentum teacher. The centre vector is updated using an exponential moving average over the teacher output. The sharpening is obtained by using a low value for the temperature $\tau_{\rm t}$ in the teacher softmax normalisation. The output probability distributions $z_t$ and $z_s$ from the teacher and the student networks are obtained as follows:
   \begin{equation}
    p_t^{(i, j)} = \frac{\exp({z_t^{(i, j)}}/{\tau_t})}{\sum_{k=1}^K \exp({z_t^{(i, k)}}/{\tau_t})},
    \end{equation}
\begin{equation}
    p_s^{(i, j)} = \frac{\exp({z_s^{(i, j)}}/{\tau_s})}{\sum_{k=1}^K \exp({z_s^{(i, k)}}/{\tau_s})}
\end{equation}

\noindent   
where $z_t$ and $z_s$ are the class logits for the student and the teacher, $p_t(i, .)$ and $p_s(i, .)$ are the probabilities corresponding to the $i-\text{th}$ token output by the teacher and the student, and $\tau_t$ and $\tau_s$ are the temperature parameters for the teacher and the student, respectively.

Our training objective is to match the output probability of the student $p_s$ with that of teacher $p_t$. We use the cross entropy measure for this task.
\begin{equation}
\mathcal{L}(\mathbf{\theta})_{\rm lcl} = \sum_k^N \left( \sum_{i=1}^n T^k_i \times \sum_{j=1}^K  - p^k_t{(i, j)} \log p^k_s{(i, j)} \right),
\end{equation}
\begin{equation}
T_{i}^k = 
\begin{cases}
    1,              & \text{if } \text{token}~ i \text{ in spectrogram } k \text{ is manipulated}\\
    0,              & \text{otherwise.}
\end{cases}
\end{equation}
 
\noindent
where $\mathbf{T}$ is a binary mask with 1 indicating the tokens corresponding to the patches that are masked or partially masked. 

The network is trained to assign a class for each token in the spectrogram and differentiate it from other tokens. In essence, the idea of local contrastive learning is similar to spectrogram reconstruction, but not in the pixel space, but rather in the class space.

As for the global contrastive leaning, we follow the same procedure as in the local contrastive learning, but proceeding only on the class token to learn a global representation for the whole spectrogram. For this particular task, two augmented views are required, where the task is to match the output representation of the class token of the student from the first augmented view, i.e. $p_{s,1}(0, .)$, to the output representation of the class token of the teacher from the second augmented view, i.e. $p_{t,2}(0, .)$, and vice versa. The overall global contrastive leaning loss is as follows: 
\begin{equation}
\begin{split}
\mathcal{L}(\mathbf{\theta})_{\rm gcl} &= \sum_k^N  \sum_{j=1}^C  - p^k_{t,1}{(0, j)} \log p^k_{s,2}{(0, j)}) \\
~~~~&+ \sum_{j=1}^C  - p^k_{t,2}{(0, j)} \log p^k_{s,1}{(0, j)}) 
\end{split}
\end{equation}

\subsection{Putting Together the ASiT Framework} 

For a given spectrogram, two augmented views are generated and passed to the teacher network. The masked version of the two views are obtained and passed to the student network. The output of the data tokens from the backbone of the student network is passed to both the reconstruction and local contrastive learning projection heads. The reconstruction loss $\mathcal{L}(\mathbf{\theta})_{\rm recons}$ is calculated between the reconstructed spectrograms and the original spectrograms. In addition, the local contrastive learning loss $\mathcal{L}(\mathbf{\theta})_{\rm lcl}$ is calculated by matching the output of the data tokens from the student network to the data tokens from the teacher network. Finally, the global contrastive learning loss $\mathcal{L}(\mathbf{\theta})_{\rm gcl}$ is calculated by matching the cross representation of the two augmented views corresponding to the class token from the student and teacher, respectively. The overall loss is then estimated as follows: 
\begin{equation}
    \mathcal{L}(\mathbf{\theta}) = \alpha_1  \mathcal{L}(\mathbf{\theta})_{\rm recons} + \alpha_2 \mathcal{L}(\mathbf{\theta})_{\rm lcl} + \alpha_3  \mathcal{L}(\mathbf{\theta})_{\rm gcl} 
    \end{equation}

    \noindent
    where $\alpha_{1, \ldots, 3}$ are the scaling factors of our multi-task objective function, which are set to 1 for simplicity. We combine the losses of the three pre-text tasks using simple averaging. We believe further improvements can be gained by optimising the weighted sum of the losses or by incorporating  the  uncertainty  weighting  approach \cite{kendall2018multi}. 

\section{Experiments}
\label{sec:exp}

We follow the common evaluation protocol to demonstrate the generalisation of the learnt features of the proposed ASiT self-supervised learning approach by pretraining the model in an unsupervised fashion on AudioSet-2M dataset, followed by finetuning on several downstream tasks, including audio event classification, keyword spotting, and speaker identification tasks.  We provide the details of the employed datasets and the implementation details of the pretraining and finetuning stages  in Section \ref{sec:datasets} and \ref{sec:imp_details}, respectively. 

\subsection{Datasets}
\label{sec:datasets}

\noindent
\textbf{AudioSet} (AS-2M, AS-20K) \cite{gemmeke2017audio} is a collection of YouTube clips for 527 multi-label weakly annotated audio events. The training set has 2 subsets; a class-wise balanced dataset with around 20K clips (AS-20K) and an unbalanced dataset combined with the balanced set, totaling approximately 2 million clips (AS-2M). The evaluation set has around 20K clips. 

\noindent
\textbf{Environmental Sound Classification} (ESC-50) \cite{piczak2015esc} is a single audio event classification dataset, consists of 2,000 5-second environmental audio recordings organized into 50 classes. We follow the standard 5-fold cross-validation to evaluate our model on this dataset.

\noindent
\textbf{Speech Commands} (SC-V2, SC-V1) \cite{warden2018speech} are two datasets for the keyword spotting task. SC-V2 consists of 105,829 1-second recordings of 35 common speech commands, split into $84,843$, $9,981$, and $11,005$ samples for the training, validation, and test set, respectively. SC-V1 is similar to SC-V2, but only contains 10 classes of keywords, 1 silence class, and 1 unknown class that includes all the other 20 common speech commands. 

\noindent
\textbf{VoxCeleb} (SID) \cite{nagrani2020voxceleb} is a dataset for speaker identification task. It contains around 150K utterances of speech from 1,251 different speakers. The dataset is split into $138,361$, $6,904$, and $8,251$ samples for the training, validation, and test set, respectively.
 
\subsection{Implementation Details}
\label{sec:imp_details}

\noindent
\textbf{Pre-training:} The backbone architecture of ASiT employs the base (ViT-B) variant of ViT \cite{dosovitskiy2020image} and is pretrained on AudioSet-2M dataset for a fair comparison with the state-of-the-art.

During the pretraining, simple data augmentation techniques are applied as we found that to learn low-level features as well as high-level semantic information, aggressive data augmentation hurts in the pretraining stage. Specifically, the given raw audio is pre-processed as a mono channel, then two random chunks of the given audio are cropped and resized to 6 seconds each. The two augmented audio waveforms are then converted into a sequence of 128-dimensional log mel filterbank (fbank) features computed with a 25ms Hanning window that shifts every 10ms. For a 6-second recording, the resulting spectrogram is of $592\times128$ dimension. The two augmented spectrograms are then normalized to adjust the statistical drifts caused by the aforementioned operations employing the mean and the standard deviation of AudioSet dataset. We conducted the pre-training under a 16,000 Hz sampling rate. For the GMML masking, the spectrograms are randomly masked  by zeros  with a total masking of 70\% of the spectrogram. 

As for the optimisation of the self-supervised pretraining, the model is trained using the Adam optimiser \cite{Loshchilov2017FixingWD} with a momentum of $0.9$ and batch size of $40$ spectrograms per GPU, using $4$ GeForce RTX 3090 GPUs. The weight decay follows a cosine schedule  from $0.04$ to $0.4$, and the base learning rate is $5e^{-4}$. The sharpening parameter of the teacher and the student are set to $\tau_{\rm t}=0.07$ and $\tau_{\rm s}=0.1$. The teacher is updated using exponential moving average of the student weights with $\lambda$ following a cosine schedule from $0.996$ to $1$ during training. 

We present the performance of ASiT under two distinct pretraining scenarios: First, ASiT undergoes pretraining for 40 epochs on the AS-2M dataset starting from scratch. Secondly, the model initially undergoes pretraining on the ImageNet-1K dataset for 400 epochs, followed by a subsequent pretraining phase on the AS-2M dataset for 40 epochs. 

\noindent
\textbf{Finetuning:}
For the downstream tasks, we discard the projection heads and perform fine-tuning using the backbone of the pretrained teacher network. Our finetuning process primarily relies on the default code provided by DeiT \cite{touvron2021training}. All models are finetuned for $60$ epochs utilising 4 GPUs with a batch size of 16 per GPU, resulting in a total batch size of 64. We employ a learning rate of $1e^{-3}$ for all datasets except AS2M, for which we use $1e^{-4}$. 

Similar to AST, we also apply frequency and time masking \cite{park2019specaugment}, random noise, and mixup \cite{tokozume2017learning} augmentation. However, it's worth noting that, unlike the pre-training phase, we refrain from recovering the introduced noise. Refer to Table \ref{tbl:finetuneRec} for more details about the hyper-parameters utilised for each dataset.

The source code and pretrained weights have been made publicly available for access\footnote{\href{https://github.com/Sara-Ahmed/ASiT}{https://github.com/ASiT}}.

\begin{table}[ht]
\caption{Finetuning hyper-parameters. BCE: Binary Cross Entropy, CE: Cross Entropy, and F/T: Frequency/Time.}
\label{tbl:finetuneRec}
\resizebox{\linewidth}{!}{
\begin{tabular}{lcccccc}
\hline
\textbf{Configuration} & \textbf{AS-2M} & \textbf{AS-20K} & \textbf{ESC-50} & \textbf{SC-V2} & \textbf{SC-V1} & \textbf{SID} \\ \hline
Weighted Sampling  & True  & False & False & False & False  & False  \\
F/T Masking \cite{park2019specaugment} & 192/48  & 192/48   & 96/24 & 48/48  & 48/48 & 192/48   \\
Mixup \cite{tokozume2017learning} & 0.5 & 0.5   & 0  & 0  & 0   & 0  \\
Loss Function & BCE  & BCE  & CE  & CE  & CE  & CE  \\
\begin{tabular}[l]{@{}l@{}}[Mean/Std] \end{tabular}  & 
        \begin{tabular}[c]{@{}c@{}}[-4.268/ \\4.569]\end{tabular}  & 
        \begin{tabular}[c]{@{}c@{}}[-4.268/ \\4.569]\end{tabular}  &
        \begin{tabular}[c]{@{}c@{}}[-6.627/ \\5.358]\end{tabular}  & 
        \begin{tabular}[c]{@{}c@{}}[-6.846/ \\5.565]\end{tabular}  &  
        \begin{tabular}[c]{@{}c@{}}[-6.702/ \\5.448]\end{tabular}  &  
        \begin{tabular}[c]{@{}c@{}}[-6.370/ \\3.074]\end{tabular}  \\
        \hline      
\end{tabular}
}
\end{table}

\begin{table*}[t]
\centering
\caption{Comparison with state-of-the-art works on audio and speech classification
tasks. Evaluation metrics are mean Average Precision (mAP) for AS-2K and accuracy (\%) for ESC-5, SC-V1, SC-V2, and SID. $^\ddagger$ with additional supervised training on AS-2M.}
\label{tbl:mainRes}
\resizebox{0.97\linewidth}{!}{
\begin{tabular}{lcccccccc}
\hline
\multirow{2}{*}{Method}           & \multirow{2}{*}{Backbone} & \multirow{2}{*}{\begin{tabular}[c]{@{}c@{}}Pretraining\\ Data\end{tabular}} & \multicolumn{5}{c}{Transfer Learning}       \\ \cline{4-9} 
 & & & AS-2M & AS-20K  & ESC-50 & SC-V2 & SC-V1 & SID  \\ \hline
\multicolumn{3}{l}{\textit{\color{gray}{{Supervised-learning-based methods}}}}\\ 
PANNs \cite{kong2020panns} & CNN   & --&  43.1 & 27.8 & 83.3 & -- & 61.8 & --  \\
AST \cite{gong2021ast}    & ViT-B & AS-2M & 45.9&28.6 & 86.8 & 96.2 & 91.6 & 35.2 \\
PaSST \cite{koutini2021efficient} & ViT-B & AS-2M & 47.1 & -- & -- & --&--&--\\
\multicolumn{3}{l}{\textit{\color{gray}{{Self-supervised-learning-based methods}}}}\\ 
COLA \cite{saeed2021contrastive} & CNN & AS-2M & -- & -- & -- & 98.1 & 95.5 & 37.7 \\
SSAST \cite{gong2022ssast}    & ViT-B & AS-2M &  -- & 29.0& 84.7 & 97.8 & 94.8 & 57.1 \\
MaskSpec \cite{chong2023masked} & ViT-B & AS-2M & 47.1&32.3  & 89.6   & 97.7    & --  & --   \\
Audio-MAE (global) \cite{xu2022masked} &ViT-B & AS-2M & 46.8&36.6 &93.6 &98.3 &97.6 &94.1\\
Audio-MAE (local) \cite{xu2022masked} &ViT-B & AS-2M & 47.3&37.0 &94.1 &98.3 &96.9 &\textbf{94.8}\\
ASiT (ours) [16kHz]  & ViT-B & AS-2M & 47.5
& {37.4} & {94.2} &{98.8}&{98.2}&85.9\\    
ASiT (ours) [16kHz]  & ViT-B & INet $\rightarrow$ AS-2M  & \textbf{48.0}
& \textbf{38.6} & \textbf{95.3} &\textbf{98.9}&\textbf{98.2}&{86.5}\\    \hline
\multicolumn{8}{l}{\textit{\color{gray}{{SSL-based methods for reference not comparison as they are pretrained on additional speech dataset LS~\cite{panayotov2015librispeech}}}}}\\  
\color{gray}{SSAST \cite{gong2022ssast}}    & 
\color{gray}{ViT-B} & \color{gray}{AS-2M + LS}   & --&
\color{gray}{31.0} & \color{gray}{88.8} & \color{gray}{98.0}  & \color{gray}{96.0}  & \color{gray}{64.3} \\
\color{gray}{MAE-AST \cite{baade2022mae}}  & \color{gray}{ViT-B} & \color{gray}{AS-2M + LS} & --&\color{gray}{30.6} & \color{gray}{90.0}  & \color{gray}{97.9} & \color{gray}{95.8}    & \color{gray}{63.3} \\ \hline
\multicolumn{8}{l}{\textit{\color{gray}{{SSL-based post arts. For reference not comparison.}}}}\\  
BEATS \cite{chen2022beats} & \color{gray}{ViT-B} & \color{gray}{AS-2M} & 
\color{gray}{48.0} &\color{gray}{38.3} &\color{gray}{95.6} &\color{gray}{98.3}&\color{gray}{97.7}  & -- \\
BEATS$^\ddagger$ \cite{chen2022beats} & \color{gray}{ViT-B} & \color{gray}{AS-2M} 
&\color{gray}{48.6} &\color{gray}{38.9} &\color{gray}{98.1} &\color{gray}{98.1} &\color{gray}{98.1} & -- \\ \hline

\end{tabular}
}
\end{table*}

\section{Results}
\label{sec:result}

We first discuss the performance and analysis of the proposed method compared to the SOTA. Further, in Section \ref{sec:ablation}, we conduct several ablation studies to investigate the effect of the individual components of the proposed approach.

\subsection{Transfer Learning}
In Table \ref{tbl:mainRes}, we compare ASiT to the supervised and self-supervised state-of-the-art approaches in audio event classification. For a fair comparison, we pretrained ASiT only on the AudioSet-2M dataset. As shown in Table \ref{tbl:mainRes}, pretrained ASiT obtains an impressive mAP of $48.0$ and $38.6$ on AudioSet-2M and AudioSet-20K, which is significantly outperforming supervised learning  with an improvement of +0.9 and +8.8 mAP and the state-of-the-art with an improvement of +1.3 and +1.6 mAP. 

Further, ASiT achieves the best performance across different audio tasks compared to other approaches. Particularly, we obtain 95.3\%, 98.9\%, and 98.2\% with an improvement of 1.3\%, 0.6\%, and 0.6\% on the ESC-50, speech command v1 and v2 datasets, respectively. 

Note that in order to improve the coverage of speech data, SSAST~\cite{gong2022ssast} and MAE-AST~\cite{baade2022mae} also used the Librispeech~\cite{panayotov2015librispeech} dataset, which has around 1,000 hours of speech. Despite pretraining ASiT only on the AudioSet-2M dataset, we outperform the aforementioned methods on the SC-V1 and SC-V2 speech tasks with a large margin, as shown in Table \ref{tbl:mainRes} and obtained a notable performance enhancement on the SID dataset, demonstrating the generalisability of the proposed self-supervised framework.

\subsection{Ablations}
\label{sec:ablation}

In all of the ablation studies, ASiT is pretrained on AS-2M from scratch for only $10$ epochs employing the small variant of vision transformers, i.e. ViT-S, as the backbone of the student and the teacher (unless mentioned otherwise). To assess the quality of the learnt representation, the pretrained models are finetuned on AS-20K and the mean average precision on the validation set is reported.

\noindent
\textbf{Effect of Different Recipes of ASiT.} The aim of this ablation study is to investigate the effect of the individual elements of the pretext learning, reported in Table \ref{tab:MCAUDIO_component}. First, we investigate the effectiveness of pretraining transformers as an autoencoder to reconstruct the input audio without any sort of masking, i.e. $D(E(\mathbf{x})) = x$, where $x$ is the input audio, $E$ is the encoder which is ViT-S in our case, and $D$ is a lightweight reconstruction decoder. Expectedly, poor performance is obtained that is slightly better than training the model from scratch. Indeed, this is attributed to the fact that without proper choice of constraints, autoencoders are capable of learning identity mapping, i.e. memorising the input without learning any useful discriminative features.

To regularise the transformer-based autoencoder, we incorporated input masking along with spectrogram reconstruction, i.e. GMML, where the mAP jumped from 11.2 to 28.4 mAP. We also investigated the effect of individually employing local contrastive learning and global contrastive learning. We found that the best individual pretext training method is GMML, followed by local contrastive learning. The least effective individual task is global contrastive learning.

Further, we investigated the effect of the different combination of the pre-text tasks. We found that using the spectrogram masking along with the reconstruction loss on its own as a means of self-supervision provided an effective starting point for efficient downstream task finetuning. Further marginal improvements can be made by extending the range of mechanisms for self-supervised pretraining. 

\begin{table}[ht]
\centering
\caption[]{Effect of the different components of ASiT for self-supervised pretraining.}
\label{tab:MCAUDIO_component}
\resizebox{\linewidth}{!}{
\begin{tabular}{x{1.25cm}x{1.3cm}x{1.3cm}x{1.3cm}x{1.2cm}}
\hline
Spectrogram Masking &  Spectrogram Reconstruction &  Local Contrastive Learning & Global Contrastive Learning & 
mAP (AS-20K) \\ \hline
\xmark & \cmark & \xmark & \xmark  & 11.2 \\
\multicolumn{5}{c}{\textit{\color{gray}{{Individual Tasks}}}}\\ 
\xmark & \xmark & \xmark & \cmark  & 17.8 \\
\cmark & \xmark & \cmark & \xmark  & 27.9\\
\cmark & \cmark & \xmark & \xmark  & 28.4 \\
\multicolumn{5}{c}{\textit{\color{gray}{{Combined Tasks}}}}\\ 
\cmark & \xmark & \cmark & \cmark  & 28.8\\
\cmark & \cmark & \xmark & \cmark  & 29.0\\  
\cmark & \cmark & \cmark & \xmark  & 29.4\\
\cmark & \cmark & \cmark & \cmark  & 31.9\\\hline 
\end{tabular}
}
\end{table}

\begin{figure}[ht]
\centering
\includegraphics[width=0.99\linewidth]{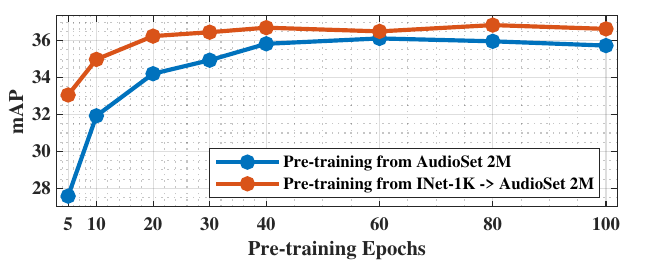}
\caption{Effect of longer pretraining.}
\label{fig:ablations_longer}
\end{figure}

\noindent
\textbf{Effect of Longer Self-supervised Pretraining.} Figure \ref{fig:ablations_longer} shows the finetuning mAP when ASiT is pretrained directly on AS-2M and when preceded by pretraining using ImageNet-1K dataset followed by AS-2M pretraining. We found that longer pretraining leads to systematic performance gain, where the mAP is steadily improving.

\begin{figure}[ht]
\includegraphics[width=0.9\linewidth]{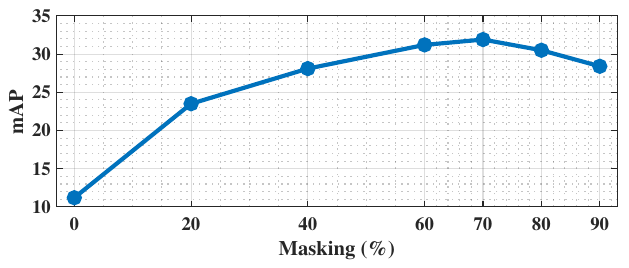}
\caption{Impact of masking percentage in pretraining.}
\label{fig:ablations_corr}
\end{figure}

\noindent
\textbf{Effect of the Percentage of Audio Masking.} Figure \ref{fig:ablations_corr} shows the mAP when the models are pretrained with different masking percentages. We found that the optimal ratio is between 60\% to 80\%. This behaviour was expected, as the masking encourages the network to learn from the unmasked regions surrounding the groups of masked regions, in order to recover the missing information. 

\begin{figure}[ht]
    \centering
    \begin{minipage}{0.32\linewidth}
    \includegraphics[width=\linewidth]{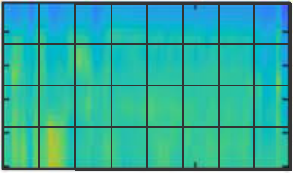}
    \caption*{\small{(a) Input}}
    \end{minipage}
    \begin{minipage}{0.32\linewidth}
    \includegraphics[width=\linewidth]{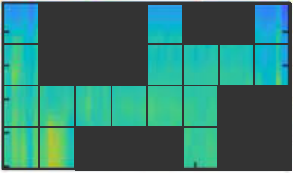}
    \caption*{(b)}
    \end{minipage}
    \begin{minipage}{0.32\linewidth}
    \includegraphics[width=\linewidth]{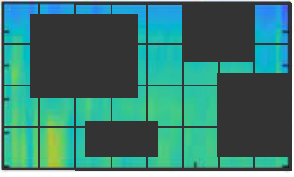}
    \caption*{(c)}
    \end{minipage}
    \caption{Example of masking the spectrogram regions while (b) aligning and (c) not aligning the masked regions with the patch boundaries.}
    \label{fig:align_withpatch}
\end{figure}

\noindent
\textbf{Effect of Aligning the Masked Regions with the Patches.} We observed that the speed of convergence and the generalisation of pretraining improve when the masking of regions is not aligned to patch boundaries. In other words, patches can be partially masked. Particularly, after 10 epochs of pretraining ASiT, the validation mAP on the AS-20K dataset is 29.2 when the masked regions are aligned, and 31.9 when the masked regions are not aligned with the patch boundaries. It is worth noting that with longer pretraining, the performance gap reduces between the two strategies. An example of masking with aligning and not aligning the masked regions with the patch boundaries is shown in Figure \ref{fig:align_withpatch}.

\noindent
\textbf{Effect of the Model Size.} In this ablation, we study the impact of model size. Specifically, we employed the small (ViT-S) and base (ViT-B) variants of vision transformer. The small variant has 12 transformer blocks with an embedding dimension of 384 and 6 attention heads. The backbone has around 22M parameters in total. As for the base variant of vision transformer, it consists of 12 transformer blocks with an embedding dimension of 768 and 12 attention heads. The backbone has around 85M parameters in total. For this ablation study, we conducted full training of both ViT-S and ViT-B models to provide a comprehensive comparison based on their respective model sizes. As shown in Table \ref{tbl:ablations_init}, the performance on the downstream task increases for the bigger model. This shows that ASiT is model size agnostic, demonstrating that it can unlock the potential of models with a higher capacity.

\begin{table}[ht]
\centering
\caption{Mean Average Precision on AS-20K pretrained with different weights' initialization and different backbones.}
\label{tbl:ablations_init}
\resizebox{0.97\linewidth}{!}{
\begin{tabular}{lcc}
\hline
\multicolumn{1}{c}{\multirow{2}{*}{\begin{tabular}[c]{@{}c@{}}Pretraining\\ Dataset\end{tabular}}} & \multicolumn{2}{c}{Backbone}              \\ \cline{2-3} 
\multicolumn{1}{c}{}     & ViT-Small& ViT-Base \\ \hline
Scratch                   &   9.1    &   11.5   \\
ImageNet-1K (INet)        &   29.7   &   32.3   \\
AudioSet-2M (AS-2M)       &   36.1   &   37.4   \\
INet $\rightarrow$ AS-2M  &   36.7   &   38.6   \\ \hline
\end{tabular}
} 
\end{table}

\noindent
\textbf{Effect of Weights Initialisation.} Despite the significant discrepancies between the image and audio modalities, the common practice for audio classification is to initialize the models from ImageNet pretrained weights. Thus, it is pertinent to answer a very important question for the audio domain: Does out-of-domain pretraining benefit audio representation learning? For that, in Table \ref{tbl:ablations_init}, we compare the performance obtained with different weight initialisation strategies across different models, including: (1) training from-scratch, i.e. no pretraining, (2) ASiT pretrained only on ImageNet-1K, (3) ASiT pretrained only on AS-2M, and (4) starting with ASiT model pretrained on ImageNet-1K, followed by pretraining on the AS-2M dataset. As shown in Table \ref{tbl:ablations_init}, the poor performance, when training from scratch, is expected, especially when the training is on small datasets, due to the lack of inductive bias in transformers. We observed that ImageNet-1K pretraining alone is not sufficient, compared to pretraining on the in-domain dataset. 

It is important to highlight that initializing with self-supervised learning pretrained weights from ImageNet-1K, followed by further pretraining on AS2M, results in enhanced performance compared to pretraining exclusively on AS2M. Nonetheless, we observed a decrease in performance when starting with supervised pretrained weights from ImageNet-1K, followed by further pretraining on AS2M.

\section{Conclusion}

We presented an efficient method of designing an audio signal classifier based on self-supervised pretraining. In our approach audio is represented by a spectrogram and processed as an image by a vision transformer. We demonstrated that the Group Masked Model Learning, that has earlier been shown to be effective for image analysis, provides also a solid basis for self-supervised pretraining for audio classification. However, its core ingredient, namely a masked autoencoder trained using reconstruction loss, has to be bolstered by other information extraction mechanisms. 

We proposed a novel transformer learning framework, constituted by a teacher-student architecture trained using a distillation approach. The training minimises a contrastive loss, and reconstruction error. As the typical global similarity learning using a classification token proved insufficient to capture discriminatory information, we proposed local similarity learning, using the alignment of local features computed for the original and masked spectrogram as a measure of similarity. 

In conclusion, the transformer pretraining, afforded by the joint use of masked spectrogram reconstruction and the combination of global/local similarity learning, was instrumental 
in obtaining very promising solutions for downstream audio classification tasks. The results of extensive evaluations of the proposed methodology on benchmarking datasets defined a new state-of-the-art performance, demonstrating the merits of the proposed methodology.

\bibliographystyle{IEEEtranN}
\bibliography{egbib}

\begin{IEEEbiography}[{\includegraphics[width=1in,height=1.25in,clip,keepaspectratio]{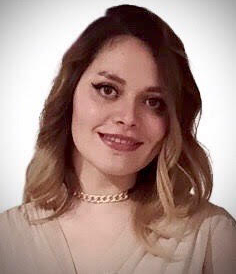}}]{Sara Atito Ali Ahmed} received her PhD in computer science from Sabanci University in 2021, with a thesis on deep learning ensembles  for image understanding. Currently, she is a Surrey Future Fellow, holding a joint position between the Centre for Vision, Speech and Signal Processing (CVSSP) and the Surrey Institute for People-Centred AI at the University of Surrey, UK. In her current role, she is working on advancing self-supervised representation learning.
\end{IEEEbiography}
 
\begin{IEEEbiography}[{\includegraphics[width=1in,height=1.25in,clip]{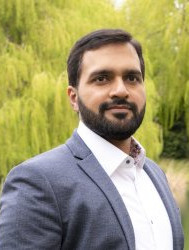}}]{Muhammad Awais}
received the B.Sc. degree in Mathematics and Physics from the AJK University in 2001, B.Sc. degree in computer engineering from UET Taxila in 2005, M.Sc. in signal processing and machine intelligence and PhD in machine learning from the University of Surrey in 2008 and 2011.
He is currently a senior lecturer in trustworthy and responsible AI at Surrey Institute for People-Centred Artificial Intelligence and Centre for Vision, Speech and Signal Processing (CVSSP).
His research interests include machine learning, deep learning, self(un,semi)-supervised learning, NLP, audio-visual analysis, medical image analysis and computer vision.
\end{IEEEbiography}

\begin{IEEEbiography}[{\includegraphics[width=1in,height=1.25in,clip]{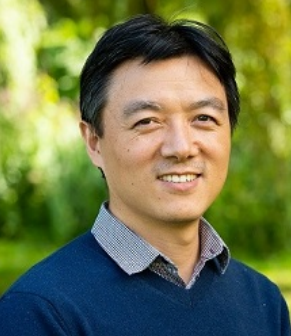}}]{Wenwu Wang} was born in Anhui, China. He received the B.Sc. degree in 1997, the M.E. degree in 2000, and the Ph.D. degree in 2002, all from Harbin Engineering University, China. He then worked in King’s College London, Cardiff University, Tao Group Ltd. (now Antix Labs Ltd.), and
Creative Labs, before joining University of Surrey, UK, in May 2007, where he is currently a professor in signal processing and machine learning, and a Co-Director of the Machine Audition Lab within the Centre for Vision Speech and Signal Processing. His current research interests include blind
signal processing, sparse signal processing, audio-visual signal processing, machine learning and perception, machine audition (listening), and statistical anomaly detection. He has (co)-authored over 300 publications in these areas. He served for IEEE Transactions on Signal Processing as an Associate Editor from 2014 to 2018, and as Senior Area Editor from 2019-2023. He is currently an Associate Editor for IEEE/ACM Transactions on Audio Speech and Language Processing. He is the elected Chair of IEEE Signal Processing Society Machine Learning for Signal Processing and elected Vice Chair of EURASIP Technical Area Committee on Audio Speech and Music Signal Processing. 
\end{IEEEbiography}

\begin{IEEEbiography}[{\includegraphics[width=1in,height=1.25in,clip]{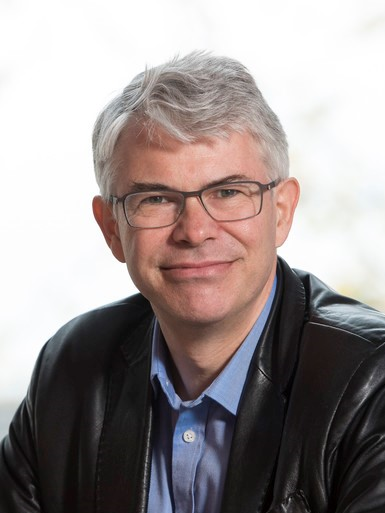}}]{Mark D Plumbley}
received the B.A.(Hons.) degree in electrical sciences
and the Ph.D. degree in neural networks from University of Cambridge, Cambridge, U.K., in 1984 and
1991, respectively. 
He is Professor of
Signal Processing at the Centre for Vision, Speech, and Signal
Processing and is head of the School of Computer Science and
Electronic Engineering at the University of Surrey, Guildford,
U.K. 
He is an expert on the analysis and processing
of audio, using a wide range of signal processing and
machine learning methods. 
He led the first international data
challenge on Detection and Classification of Acoustic Scenes
and Events, and currently holds an Engineering and Physical
Sciences Research Council Fellowship for ``AI for Sound,'' on
the automatic recognition of everyday sounds. 
He is a Member of the IEEE Signal Processing Society Technical Committee on 
Audio and Acoustic Signal Processing, and a Fellow of the IET and IEEE.
\end{IEEEbiography}

\begin{IEEEbiography}[{\includegraphics[width=1in,height=1.25in,clip,keepaspectratio]{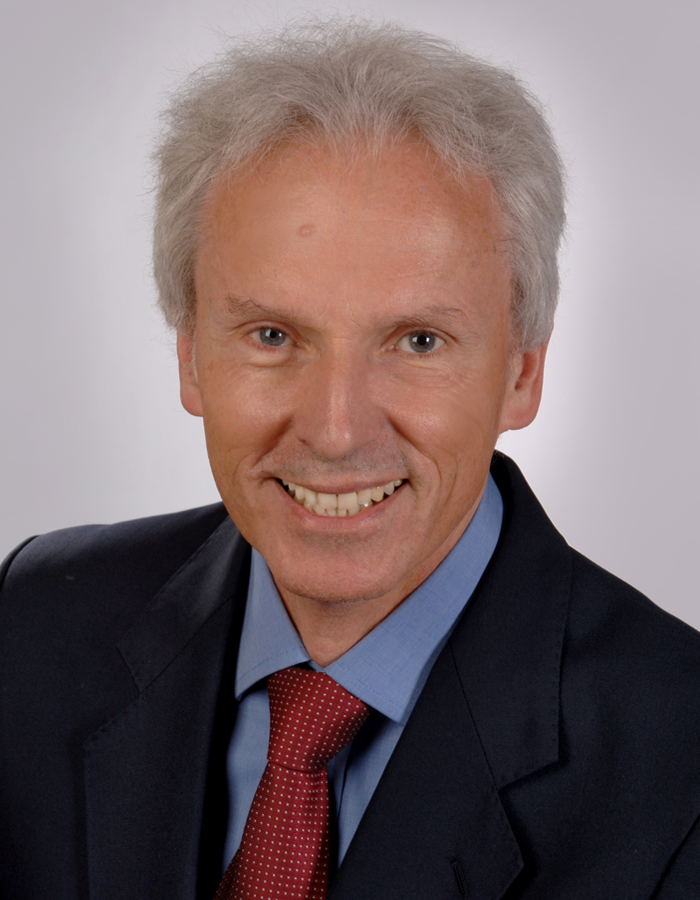}}]{Josef Kittler}
(M'74-LM'12) received the B.A., Ph.D., and D.Sc. degrees from the University of Cambridge, in 1971, 1974, and 1991, respectively.
He is a distinguished Professor of Machine Intelligence at the Centre for Vision, Speech and Signal Processing, University of Surrey, Guildford, U.K.
He conducts research in biometrics, video and image dataset retrieval, medical image analysis, and
cognitive vision. He published the textbook Pattern Recognition: A Statistical Approach and over 700 scientific papers. 
His publications have been cited more than 68,000 times (Google Scholar).

 He currently serves on the Editorial Boards of Pattern Recognition Letters and  Pattern Recognition. He also served as a member of the Editorial Board of IEEE Transactions on Pattern Analysis and Machine Intelligence during 1982-1985. He served on the Governing Board of the International Association for Pattern Recognition (IAPR) as one of the two British representatives during the period 1982-2005, President of the IAPR during 1994-1996.
\end{IEEEbiography}

\vfill

\end{document}